\documentclass[trackchanges,twocolumn]{aastex701}
\usepackage{longtable} 
\usepackage{booktabs}

\shorttitle{AT2025ulz and S250818k: Leveraging DESI spectroscopy}
\shortauthors{Hall et al.}

\newcommand{\mcwilliams}{
    McWilliams Center for Cosmology and Astrophysics,
    Department of Physics,
    Carnegie Mellon University,
    5000 Forbes Avenue, Pittsburgh, PA 15213, USA
}
\newcommand{\lmu}{
    University Observatory, 
    Faculty of Physics, 
    Ludwig-Maximilians-Universität München, 
    Scheinerstr. 1, 81679 Munich, Germany
}

\begin{document}

\title{AT2025ulz and S250818k: Leveraging DESI spectroscopy in the hunt for a kilonova associated with a sub-solar mass gravitational wave candidate}

\author[0000-0002-9364-5419]{Xander J. Hall}
\affiliation{\mcwilliams}
\email[show]{xhall@cmu.edu}

\author[0000-0002-6011-0530]{Antonella Palmese}
    \affiliation{\mcwilliams}
    \email{apalmese@andrew.cmu.edu}
\author[0000-0002-9700-0036]{Brendan O'Connor}
    \altaffiliation{McWilliams Fellow}
    \affiliation{\mcwilliams}
    \email{boconno2@andrew.cmu.edu}

\author[0000-0003-3270-7644]{Daniel Gruen}
\affiliation{\lmu}
\affiliation{Excellence Cluster ORIGINS, Boltzmannstr. 2, 85748 Garching, Germany}
\email{daniel.gruen@lmu.de}

\author[0009-0001-0574-2332]{Malte Busmann}
\affiliation{\lmu}
\email{m.busmann@physik.lmu.de}

\author[0009-0008-2754-1946]{Julius Gassert}
\affiliation{\lmu}
\affiliation{\mcwilliams}
\email{julius.gassert@campus.lmu.de}

\author[0000-0001-7201-1938]{Lei Hu}
\affiliation{\mcwilliams}
\email{leihu@andrew.cmu.edu}

\author[0000-0003-2362-0459]{Ignacio Maga\~na~Hernandez}
\affiliation{\mcwilliams}
\email{imhernan@andrew.cmu.edu}

\author{Jessica Nicole Aguilar}
\affiliation{Lawrence Berkeley National Laboratory, 1 Cyclotron Road, Berkeley, CA 94720, USA}
\email{jaguilar@lbl.gov}

\author[0000-0003-3433-2698]{Ariel Amsellem}
\affiliation{\mcwilliams}
\email{aamselle@andrew.cmu.edu}

\author[0000-0001-6098-7247]{Steven Ahlen}
\affiliation{Department of Physics, Boston University, 590 Commonwealth Avenue, Boston, MA 02215 USA}
\email{ahlen@bu.edu}

\author[0000-0003-0776-8859]{John Banovetz}
\affiliation{Lawrence Berkeley National Laboratory, 1 Cyclotron Road, Berkeley, CA 94720, USA}
\email{jdbanovetz@lbl.gov}

\author[0000-0001-5537-4710]{Segev BenZvi}
\affiliation{Department of Physics \& Astronomy, University of Rochester, 206 Bausch and Lomb Hall, P.O. Box 270171, Rochester, NY 14627-0171, USA}
\email{sbenzvi@ur.rochester.edu}

\author[0000-0001-9712-0006]{Davide Bianchi}
\affiliation{Dipartimento di Fisica ``Aldo Pontremoli'', Universit\`a degli Studi di Milano, Via Celoria 16, I-20133 Milano, Italy}
\affiliation{INAF-Osservatorio Astronomico di Brera, Via Brera 28, 20122 Milano, Italy}
\email{davide.bianchi1@unimi.it}

\author{David Brooks}
\affiliation{Department of Physics \& Astronomy, University College London, Gower Street, London, WC1E 6BT, UK}
\email{david.brooks@ucl.ac.uk}

\author[0000-0001-7316-4573]{Francisco Javier Castander}
\affiliation{Institut d'Estudis Espacials de Catalunya (IEEC),  Esteve Terradas 1, Edifici RDIT, Campus PMT-UPC, 08860 Castelldefels, Spain
}
\affiliation{Institute of Space Sciences, ICE-CSIC, Campus UAB, Carrer de Can Magrans, 08913 Bellaterra, Barcelona, Spain
}
\email{fjc@ice.csic.es}

\author{Todd Claybaugh}
\affiliation{Lawrence Berkeley National Laboratory, 1 Cyclotron Road, Berkeley, CA 94720, USA}
\email{tmclaybaugh@lbl.gov}

\author[0000-0002-2169-0595]{Andrei Cuceu}
\affiliation{Lawrence Berkeley National Laboratory, 1 Cyclotron Road, Berkeley, CA 94720, USA}
\email{acuceu@lbl.gov}

\author[0000-0002-4928-4003]{Arjun Dey}
\affiliation{NSF NOIRLab, 950 N. Cherry Ave., Tucson, AZ 85719, USA}
\email{arjun.dey@noirlab.edu}

\author{Peter Doel}
\affiliation{Department of Physics \& Astronomy, University College London, Gower Street, London, WC1E 6BT, UK}
\email{apd@star.ucl.ac.uk}

\author[0009-0006-7670-9843]{Jennifer Fabà-Moreno}
\affiliation{\lmu}
\email{J.Faba@campus.lmu.de}

\author[0000-0003-4992-7854]{Simone Ferraro}
\affiliation{Lawrence Berkeley National Laboratory, 1 Cyclotron Road, Berkeley, CA 94720, USA}
\affiliation{University of California, Berkeley, 110 Sproul Hall \#5800 Berkeley, CA 94720, USA}
\email{sferraro@lbl.gov}

\author[0000-0002-3033-7312]{Andreu Font-Ribera}
\affiliation{Institut de F\'{i}sica d’Altes Energies (IFAE), The Barcelona Institute of Science and Technology, Edifici Cn, Campus UAB, 08193, Bellaterra (Barcelona), Spain}
\email{afont@ifae.es}

\author[0000-0002-2890-3725]{Jaime E. Forero-Romero}
\affiliation{Departamento de F\'isica, Universidad de los Andes, Cra. 1 No. 18A-10, Edificio Ip, CP 111711, Bogot\'a, Colombia}
\email{je.forero@uniandes.edu.co}

\author{Gaston Gutierrez}
\affiliation{Fermi National Accelerator Laboratory, PO Box 500, Batavia, IL 60510, USA}
\email{gaston@fnal.gov}

\author[0000-0001-7178-8868]{Laurent Le Guillou}
\affiliation{Sorbonne Universit\'{e}, CNRS/IN2P3, Laboratoire de Physique Nucl\'{e}aire et de Hautes Energies (LPNHE), FR-75005 Paris, France}
\email{llg@lpnhe.in2p3.fr}

\author[0000-0003-0201-5241]{Dick Joyce}
\affiliation{NSF NOIRLab, 950 N. Cherry Ave., Tucson, AZ 85719, USA}
\email{richard.joyce@noirlab.edu}

\author[0000-0003-3510-7134]{Theodore Kisner}
\affiliation{Lawrence Berkeley National Laboratory, 1 Cyclotron Road, Berkeley, CA 94720, USA}
\email{tskisner@lbl.gov}

\author[0000-0001-6356-7424]{Anthony Kremin}
\affiliation{Lawrence Berkeley National Laboratory, 1 Cyclotron Road, Berkeley, CA 94720, USA}
\email{akremin@lbl.gov}

\author[]{Ofer Lahav}
\affiliation{Department of Physics \& Astronomy, University College London, Gower Street, London, WC1E 6BT, UK}
\email{o.lahav@ucl.ac.uk}

\author[0000-0002-6731-9329]{Claire Lamman}
\affiliation{The Ohio State University, Columbus, 43210 OH, USA}
\email{lamman.1@osu.edu}

\author[0000-0003-1838-8528]{Martin Landriau}
\affiliation{Lawrence Berkeley National Laboratory, 1 Cyclotron Road, Berkeley, CA 94720, USA}
\email{mlandriau@lbl.gov}

\author[0000-0003-1887-1018]{Michael Levi}
\affiliation{Lawrence Berkeley National Laboratory, 1 Cyclotron Road, Berkeley, CA 94720, USA}
\email{melevi@lbl.gov}

\author[0000-0002-1769-1640]{Axel de la Macorra}
\affiliation{Instituto de F\'{\i}sica, Universidad Nacional Aut\'{o}noma de M\'{e}xico,  Circuito de la Investigaci\'{o}n Cient\'{\i}fica, Ciudad Universitaria, Cd. de M\'{e}xico  C.~P.~04510,  M\'{e}xico}
\email{macorra@fisica.unam.mx}

\author[0000-0003-4962-8934]{Marc Manera}
\affiliation{Departament de F\'{i}sica, Serra H\'{u}nter, Universitat Aut\`{o}noma de Barcelona, 08193 Bellaterra (Barcelona), Spain}
\affiliation{Institut de F\'{i}sica d’Altes Energies (IFAE), The Barcelona Institute of Science and Technology, Edifici Cn, Campus UAB, 08193, Bellaterra (Barcelona), Spain}
\email{mmanera@ifae.es}

\author[0000-0002-1125-7384]{Aaron Meisner}
\affiliation{NSF NOIRLab, 950 N. Cherry Ave., Tucson, AZ 85719, USA}
\email{aaron.meisner@noirlab.edu}

\author{Ramon Miquel}
\affiliation{Instituci\'{o} Catalana de Recerca i Estudis Avan\c{c}ats, Passeig de Llu\'{\i}s Companys, 23, 08010 Barcelona, Spain}
\affiliation{Institut de F\'{i}sica d’Altes Energies (IFAE), The Barcelona Institute of Science and Technology, Edifici Cn, Campus UAB, 08193, Bellaterra (Barcelona), Spain}
\email{rmiquel@ifae.es}

\author[0000-0002-2733-4559]{John Moustakas}
\affiliation{Department of Physics and Astronomy, Siena University, 515 Loudon Road, Loudonville, NY 12211, USA}
\email{jmoustakas@siena.edu}

\author[0000-0001-9070-3102]{Seshadri Nadathur}
\affiliation{Institute of Cosmology and Gravitation, University of Portsmouth, Dennis Sciama Building, Portsmouth, PO1 3FX, UK}
\email{seshadri.nadathur@port.ac.uk}

\author[0000-0001-7145-8674]{Francisco Prada}
\affiliation{Instituto de Astrof\'{i}sica de Andaluc\'{i}a (CSIC), Glorieta de la Astronom\'{i}a, , E-18008 Granada, Spain}
\email{fprada@iaa.es}

\author[0000-0001-6979-0125]{Ignasi Pérez-Ràfols}
\affiliation{Departament de F\'isica, EEBE, Universitat Polit\`ecnica de Catalunya, Eduard Maristany 10, 08930 Barcelona, Spain}
\email{ignasi.perez.rafols@upc.edu}

\author{Graziano Rossi}
\affiliation{Department of Physics and Astronomy, Sejong University, 209 Neungdong-ro, Gwangjin-gu, Seoul 05006, Republic of Korea}
\email{graziano@sejong.ac.kr}

\author[0000-0002-9646-8198]{Eusebio Sanchez}
\affiliation{CIEMAT, Avenida Complutense 40, E-28040 Madrid, Spain}
\email{eusebio.sanchez@ciemat.es}

\author{David Schlegel}
\affiliation{Lawrence Berkeley National Laboratory, 1 Cyclotron Road, Berkeley, CA 94720, USA}
\email{djschlegel@lbl.gov}

\author{Michael Schubnell}
\affiliation{Department of Physics, University of Michigan, 450 Church Street, Ann Arbor, MI 48109, USA}
\email{schubnel@umich.edu}

\author{David Sprayberry}
\affiliation{NSF NOIRLab, 950 N. Cherry Ave., Tucson, AZ 85719, USA}
\email{david.sprayberry@noirlab.edu}

\author[0000-0003-1704-0781]{Gregory Tarlé}
\affiliation{University of Michigan, 500 S. State Street, Ann Arbor, MI 48109, USA}
\email{gtarle@umich.edu}

\author{Benjamin Alan Weaver}
\affiliation{NSF NOIRLab, 950 N. Cherry Ave., Tucson, AZ 85719, USA}
\email{benjamin.weaver@noirlab.edu}

\author[0000-0001-5381-4372]{Rongpu Zhou}
\affiliation{Lawrence Berkeley National Laboratory, 1 Cyclotron Road, Berkeley, CA 94720, USA}
\email{rongpuzhou@lbl.gov}

\author[0000-0002-6684-3997]{Hu Zou}
\affiliation{National Astronomical Observatories, Chinese Academy of Sciences, A20 Datun Road, Chaoyang District, Beijing, 100101, P.~R.~China}
\email{zouhu@nao.cas.cn}

\collaboration{all}{The Dark Energy Spectroscopic Instrument Collaboration}

\begin{abstract}

On August 18th, 2025, the LIGO–Virgo–KAGRA collaboration reported a sub-threshold gravitational wave candidate detection consistent with a sub-solar-mass neutron star merger, denoted S250818k. An optical transient, AT2025ulz, was discovered within the localization region. AT2025ulz initially appeared to meet the expected behavior of kilonova (KN) emission, the telltale signature of a binary neutron star merger.  
The transient subsequently rebrightened after $\sim$\,$5$ days and developed spectral features characteristic of a Type IIb supernova. In this work, we analyze the observations of the host galaxy of AT2025ulz obtained by the Dark Energy Spectroscopic Instrument (DESI). From the DESI spectrum, we obtain a secure redshift of $z = 0.084840 \pm 0.000006$, which places the transient within $2\sigma$ of the gravitational wave distance and results in an integral overlap between the gravitational wave alert and the transient location of $\log_{10}\mathcal{I} \approx 3.9-4.2$. Our analysis of the host galaxy's spectral energy distribution reveals a star-forming, dusty galaxy with stellar mass ${\sim} 10^{10}~M_\odot$, broadly consistent with the population of both short gamma-ray bursts and core-collapse supernova host galaxies. We also present our follow-up of DESI-selected candidate host galaxies using the Fraunhofer Telescope at the Wendelstein Observatory, and show the promise of DESI for associating or rejecting candidate electromagnetic counterparts to gravitational wave alerts. These results emphasize the value of DESI’s extensive spectroscopic dataset in rapidly characterizing host galaxies, enabling spectroscopic host subtraction, and guiding targeted follow-up. 

\end{abstract}

\keywords{\uat{Galaxies}{573} --- \uat{Time domain astronomy}{2109} --- \uat{Gravitational waves}{678} --- \uat{Transient sources}{1851} --- \uat{Type II supernovae}{1731} 
}


\section{Introduction}
\label{intro} 

The multi-messenger astronomy community has been rigorously chasing potential electromagnetic counterparts to gravitational wave events for over a decade. The watershed discovery of the binary neutron star merger gravitational wave signature GW170817 \citep{abbott_bns_2017} with its associated short gamma-ray burst \citep{Goldstein2017,Savchenko2017} and kilonova \citep[e.g.,][]{Coulter2017,Troja2017,Evans2017,Arcavi2017,SoaresSantos2017,Drout2017} provided critical insights into compact binary mergers. Despite significant follow-up efforts and some candidate counterparts reports with uncertain associations, it has been 8 years since the discovery of GW170817 and the community has yet to confidently associate a second electromagnetic counterpart to a gravitational wave source, including in the ongoing fourth observing run (O4) of the LIGO-Virgo-KAGRA (LVK) collaboration. Based on the dearth of high-significance binary neutron star merger detections since GW170817 and GW190425, the inferred rate of binary neutron star mergers has steadily decreased with each subsequent LVK observing run \citep{abbott_population_2023, the_ligo_scientific_collaboration_gwtc-40_2025}. 

On August 18th, 2025, a low-significance gravitational wave candidate S250818k was detected by LVK \citep{2025GCN.41437....1L}. The signal was estimated to have a probability of being a terrestrial artifact of 71\% and a 29\% chance to result from a binary neutron star merger \citep{2025GCN.41437....1L}. Assuming the event is of astrophysical origin, there is a $>$\,$99\%$ probability that the event involved at least one object consistent in mass with a neutron star, and that it disrupted material outside of the remnant object\footnote{\url{https://emfollow.docs.ligo.org/userguide/content.html}}. Moreover, given that the chirp mass is $<0.87~M_\odot$ \citep{2025GCN.41437....1L}, there is significant probability that at least one of the two objects has sub-solar mass, rendering this source particularly interesting. The Zwicky Transient Facility (ZTF) followed up part of the sky localization of S250818k and identified a candidate optical counterpart, AT2025ulz \citep{GCN41414}. The candidate counterpart subsequently rapidly faded and reddened over the first few days \citep{GCN41421,GCN41433,GCN41480}, in a fashion typical of a kilonova \citep{Kasliwal2025sn,Hall2025sn}. The identification of the source as a kilonova is called into question by its short lived decay, which plateaued after a few days. The subsequent rise occurred in the fashion of a supernova \citep{2025GCN.41507....1F,2025GCN.41532....1B,2025GCN.41544....1B,2025GCN.41540....1G}. In fact, spectroscopic features typical of IIb supernovae were quickly identified during this rising phase \citep{2025GCN.41532....1B,Gillanders25ulz,Yang2025ulz,Franz2025,Kasliwal2025sn}. Regardless, AT2025ulz is an interesting transient with a tantalizing possible connection to S250818k. The analysis of the extensive imaging and spectroscopic dataset \citep{Kasliwal2025sn} is complicated by the presence of a luminous host galaxy directly underlying the transient's position and the limited availability of pre-explosion templates. 

Meanwhile, The Dark Energy Spectroscopic Instrument (DESI) has been consistently building the largest spectroscopic catalog of galaxies to date, recently surpassing over 40 million spectra \citep{schlafly_survey_2023, desi_collaboration_data_2025, desi_collaboration_desi_2025}. Such a catalog provides novel opportunities in transient science (e.g., The DESI Transient Survey, \citealt{2025TNSAN..84....1H, Hall2025dts}, the MOST Hosts Survey \citealt{2024ApJS..275...22S}). DESI offers the ability to quickly identify and eliminate or highlight candidate counterparts that either fall within or outside the volumetric region of an LVK alert. One of the most difficult aspects of real-time transient science is the separation of transient light from their hosts, especially in the case of fainter transients. While imaging templates from wide-field surveys generally exist over most of the sky in optical filters, the availability of spectroscopic templates has been substantially lacking. Long-slit spectroscopy can be contaminated by host galaxy light, depending on the position of the transient with respect to its host and their relative brightnesses. This impacts the interpretation of transient spectra (such as blackbody temperature), identification of absorption features, and the derivation of the intrinsic line fluxes and ratios of the underlying galaxy. Pre-transient DESI spectra provide an unparalleled ability to rapidly perform spectroscopic host subtraction. AT2025ulz serendipitously occurred in one of these 40 million DESI spectra. In this \emph{Letter}, we model the host galaxy using both photometry and spectroscopy, analyze its properties and compare them to other transients, estimate the probability of association with the GW alert, present our follow-up observations around DESI galaxies, and show the power of a high-completeness spectroscopic survey such as DESI for multi-messenger astronomy.

Throughout this paper, we report all magnitudes in the AB system, and we assume a Planck 2018 \citep{aghanim_planck_2020} cosmology with the following parameters: $H_{0} = 67.66~\mathrm{km\,s^{-1}\,Mpc^{-1}}$,  $\Omega_{m} = 0.30966$, $T_{\mathrm{CMB}} = 2.7255~\mathrm{K}$, $N_{\mathrm{eff}} = 3.046$, $m_{\nu} = [0.0,\, 0.0,\, 0.06]~\mathrm{eV}$, and $\Omega_{b} = 0.04897$.

\begin{figure}
    \centering
    \includegraphics[width=1\linewidth]{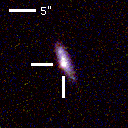}
    \caption{False-color image in $grz$ bands of AT2025ulz from the first epoch of Wendelstein Observatory imaging on August 18, 2025 \citep{GCN41421,Hall2025sn}. The transient is visibly off-center from the galaxy. Star forming regions of the galaxy are noticeable as bluer patches. In this color image the transient is visually enhanced by co-adding with a difference image with respect to Legacy Survey data \citep{Dey2019AJ}.}
    \label{fig:finder}
\end{figure}

\section{Dark Energy Spectroscopic Instrument}

\begin{figure*}
    \includegraphics[width = 0.56\linewidth]{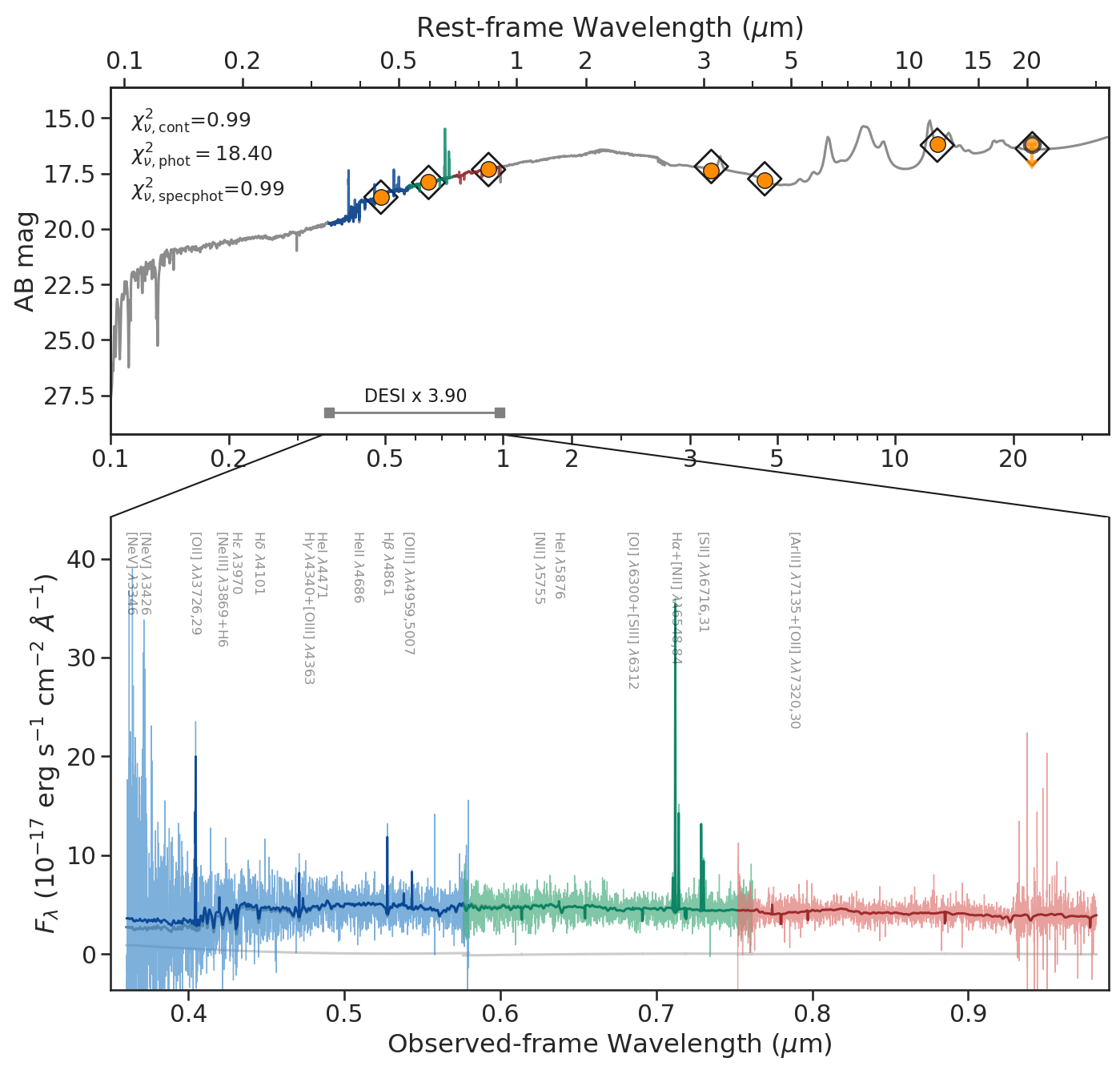}
    \includegraphics[width = 0.44\linewidth]{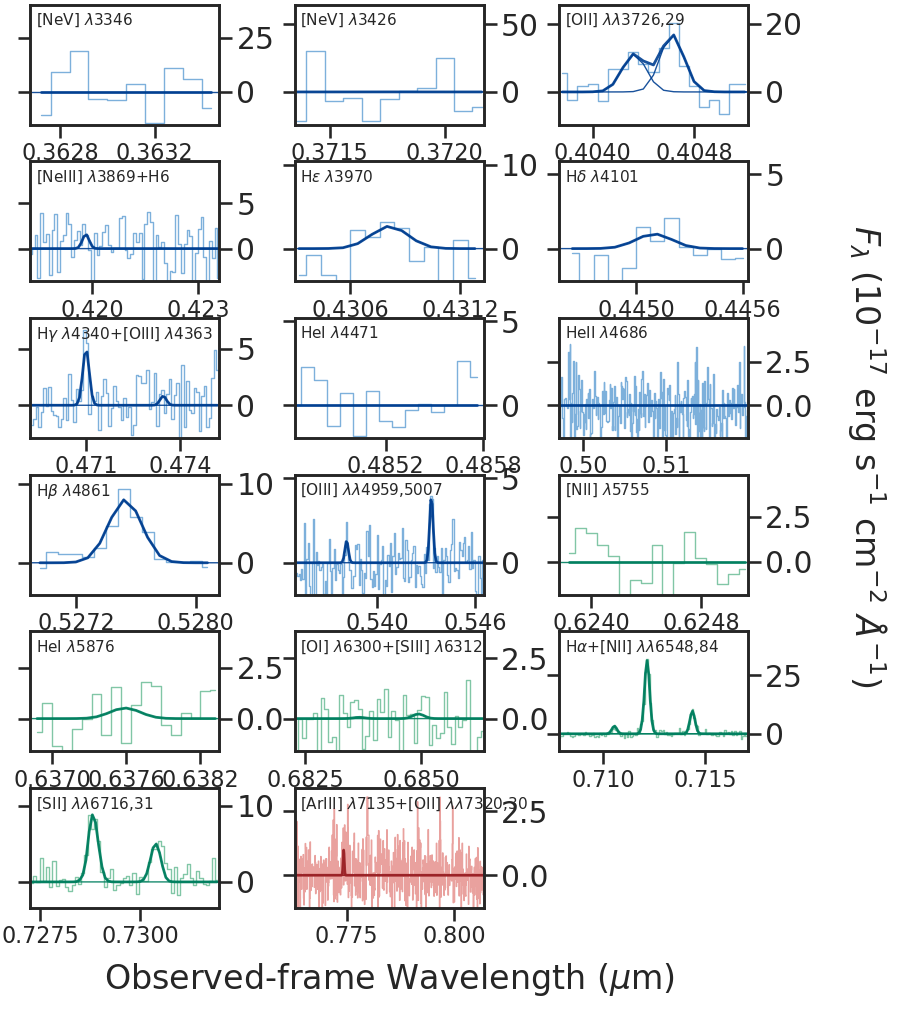}
    \caption{\emph{Left:} Spectral energy distribution fit from \texttt{FastSpecFit}, combining Legacy Survey and WISE photometry with DESI spectra. The lower panel shows the stellar population model spectrum, with key emission lines labeled.  
\emph{Right:} Gaussian fits to individual spectral features used to measure line properties. The dark curve shows the best-fit model, while the faint lines represent the DESI data. We derive a redshift $z=0.084840 \pm 0.000006$.}
    \label{fig:fastspecfit}
\end{figure*}

DESI is comprised of 5,000 fibers that can each be positioned independently on sky \citep{poppett_overview_2024}. Each fiber is 107 $\mu$m in diameter, which is equivalent to $\sim1.5"$ on the sky. The spectrograph utilizes three cameras in the approximate $BRZ$ bands that cover $3600-9824$ \AA\ at a spectral resolution $R \sim 2000 - 5500$ \citep{desi_collaboration_desi_2016, miller_optical_2024, desi_collaboration_overview_2022, desi_collaboration_data_2025, desi_collaboration_desi_2025}. Each spectrum is reduced and the flux calibrated with the DESI spectroscopic data pipeline \citep{guy_spectroscopic_2023}. To establish a redshift from a given DESI spectrum, a decomposition into a combination of template spectra is performed using Redrock\footnote{\url{https://github.com/desihub/redrock/}} and a redshift is determined based on minimizing $\chi^2$ \citep[see][]{Redrock.Bailey.2024}. As part of DESI's Bright Galaxy Survey \citep{Hahn2023}, the host galaxy of AT2025ulz was observed on 2023-03-19 and is part of the DESI data release 2 \citep[DR2][]{desi_collaboration_data_2025, desi_collaboration_desi_2025}. We derive a redshift $z=0.084840 \pm 0.000006$ (see Figure~\ref{fig:fastspecfit}). 

\section{Spectral Analysis} \label{sec:anal}

\subsection{Fastspecfit}

\begin{figure*}
    \includegraphics[width = 0.98\linewidth]{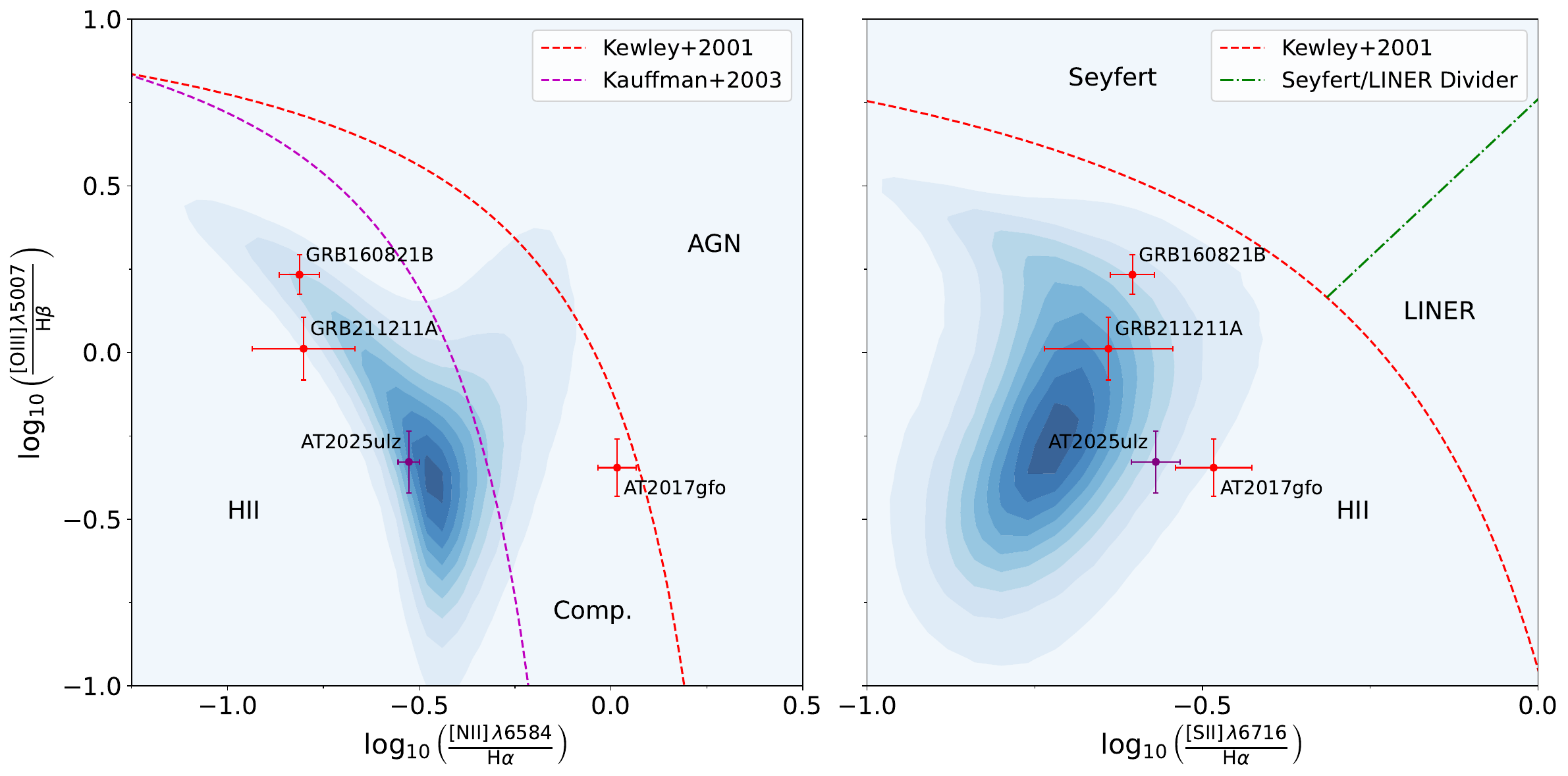}
    \caption{A BPT \citep{baldwin_classification_1981} analysis of the host of AT2025ulz (\textcolor{violet}{violet}) based on its  DESI spectrum and including for comparison the host galaxies of known gamma ray bursts (GRB) with an associated candidate KN (GRBs 160821B and 211211A) observed by DESI. The galaxy is a standard star forming galaxy, similar to the hosts of known GRB-KN. Notably, the host of AT2017gfo (associated to GW170817) has weak AGN lines that are not present in any of the other galaxies \citep{palmese_evidence_2017}. The background contours are made from the \texttt{FastSpecFit} value added catalog for DESI DR2 data \texttt{FastSpecFit} (version 3.2.0; \citealt{2023ascl.soft08005M, desi_collaboration_desi_2025, Moustakas20XXfastspecfit}). Separation lines are drawn from \citet{kewley_theoretical_2001}, \citet{kauffmann_host_2003}, and \citet{kewley_host_2006}.}
    \label{fig:bpt}
\end{figure*}

We measure the rest-frame optical emission-line fluxes, equivalent widths, and line-ratios of the AT 2025ulz host galaxy using \texttt{FastSpecFit} (version 3.2.0; \citealt{2023ascl.soft08005M, Moustakas20XXfastspecfit}; see Fig. \ref{fig:fastspecfit}). FastSpecFit performs a joint (simultaneous) fit of the DESI three-camera optical spectrophotometry and the broadband optical plus infrared photometry from the DESI Legacy Imaging Surveys Data Release 9 \citep{dey_overview_2019}. We model the stellar continuum as a non-negative least-squares sum of five solar-metallicity, logarithmically spaced age bins between 15~Myr and 12.7~Gyr and a $\tau\propto\lambda^{-0.7}$ dust law. We then subtract the stellar continuum model from the DESI spectrum and measure all the emission lines in the observed spectral range as using constrained Gaussian line-profiles. We estimate the uncertainties on the continuum and emission-line parameters using 1000 Monte Carlo realizations of the data for AT2025ulz and 50 for the other sGRBs. One noteworthy component of AT2017gfo was the AGN activity of the host \citep{palmese_evidence_2017}. To determine if such weak AGN lines are present in the host of AT2025ulz, we perform a Baldwin, Phillips \& Terlevich \citep[BPT;][]{baldwin_classification_1981} analysis (Figure~\ref{fig:bpt}). We determine that the emission lines of are more consistent with that of a purely star forming galaxy, similar to known GRB-KN also in DESI.

\begin{table}
\centering
\scriptsize
\caption{Emission line measurements and ratios for the AT~2025ulz host galaxy based on the \texttt{FastSpecFit} run. The $A_v$ extinction is computed from the Balmer decrement with the Calzetti Milky Way dust law with $R_V=3.1$ \citep{calzetti_dust_1994}. We compute the SFR using Equation 2 from \citet{murphy_calibrating_2011}}
\begin{tabular}{lcc}
\hline\hline
Quantity & Value\\
\hline
$z$ & $0.084840 \pm 0.000006$\\ 
EW(H$\alpha$) [\AA] & $22.72 \pm 0.66$\\
EW([OII]) [\AA] & $19.6 \pm 3.75$\\
EW([OIII]) [\AA] & $2.14 \pm 0.30$\\
H$\alpha$/H$\beta$ & $4.67 \pm 0.48$\\
$\log_{10}$([OIII]$\lambda5007$/H$\beta$) & $-0.33 \pm 0.07$ \\
$\log_{10}$([NII] $\lambda6584$/H$\alpha$) & $-0.53 \pm 0.03$\\
$\log_{10}$([SII] $\lambda6716$/H$\alpha$) & $-0.57 \pm 0.04$\\
{[OII]} $\lambda3726/\lambda3729$ & $0.66 \pm 0.18$\\
{[SII]} $\lambda6731/\lambda6716$ & $0.57 \pm 0.10$\\
\hline
$A_V$ [mag] & $1.302 \pm 0.272$\\
$\mathrm{SFR}$ [$M_\odot$/yr] & $0.386 \pm 0.104$\\
\hline
\label{tab:fastspecfit_output}
\end{tabular}
\end{table}

\subsection{Prospector}

The complete spectral energy distribution was also modeled with \texttt{prospector} \citep{2017ApJ...837..170L,Johnson2019,2021ApJS..254...22J}. For this fit, we use photometry based on the \citet{Kron1980} radius produced by \texttt{HostPhot} \citep{muller-bravo_hostphot_2022}. We derive a Kron radius from $grz$ imaging from Legacy Survey \citep{dey_overview_2019} and applied that aperture to FUV and NUV imaging from GALEX \citep{bianchi_revised_2017}, $ugriz$ imaging from the Sloan Digital Sky Survey \citep[SDSS;][]{ahumada_16th_2020}, $grizy_{P1}$ imaging from PanSTARRS \citep{2010SPIE.7733E..0EK, chambers_pan-starrs1_2016}, $J$, $H$, and $K_s$ imaging from the Two Micron All-Sky Survey \citep[2MASS;][]{Skrutskie2006}. Due to the vast pixel difference in mid-IR imaging, we fit a new Kron radius for W1 and apply it to all imaging from the Wide-field Infrared Explorer \citep[WISE;][]{wright_wide-field_2010}. All photometry is corrected for Milky Way extinction using dust maps from \citet{schlafly_measuring_2011} and the extinction curve from \citet{fitzpatrick_correcting_1999}.

\begin{figure*}
    
    \begin{center}
    \includegraphics[width = \linewidth]{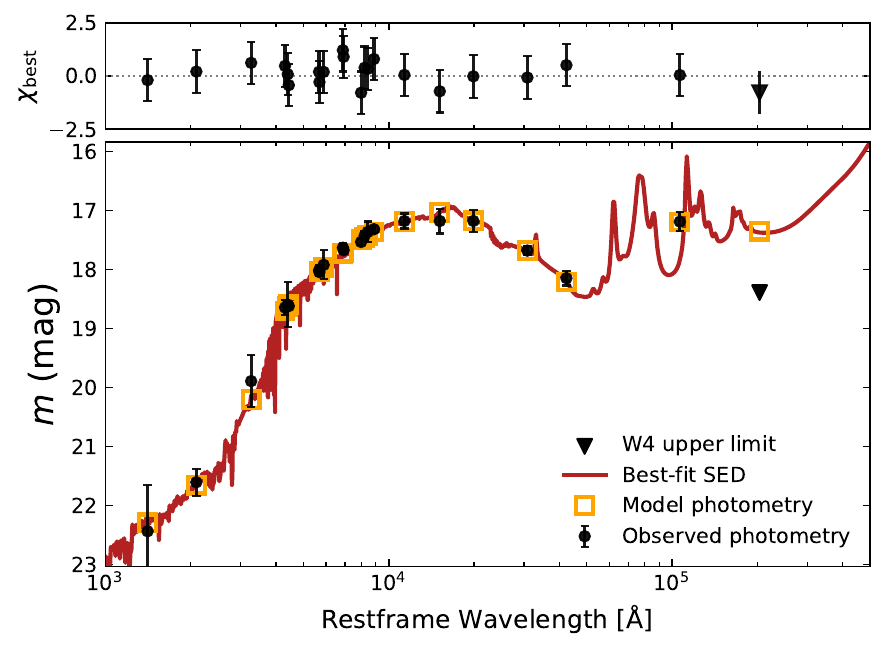}
    \caption{The best-fit host galaxy SED derived using \texttt{prospector}. \citep{Johnson2019} Model photometry is shown as orange squares and compared to the observed photometry (black circles). The top panel shows the residuals compared to the best-fit SED (red line). }
    \label{fig:prospector_plot_SED}
        \end{center}
\end{figure*}

\begin{figure*}
    
    \begin{center}
    \includegraphics[width = \linewidth]{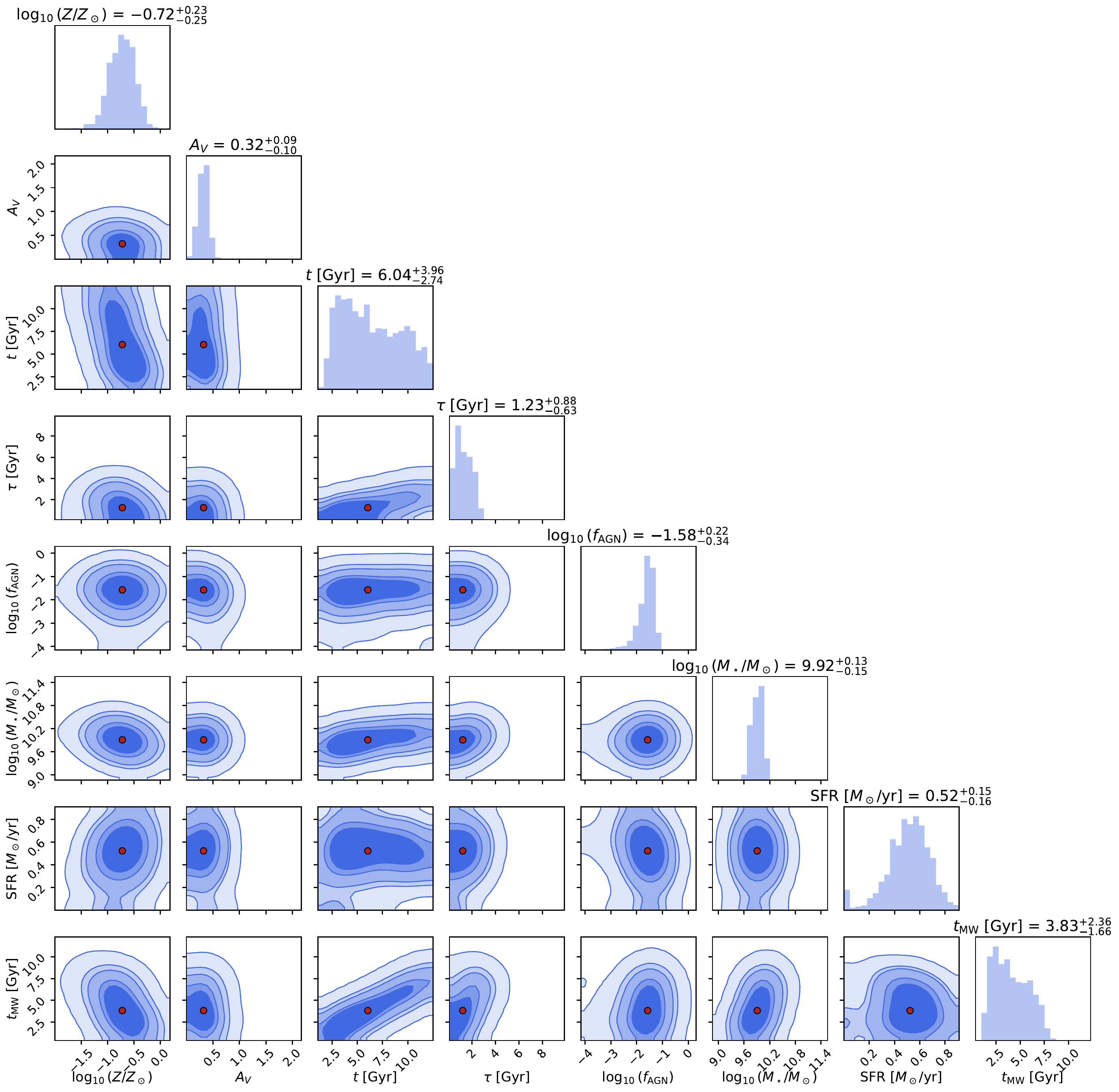}
    \caption{Corner plot showing the posterior distribution of the host galaxy parameters derived using \texttt{prospector} \citep{Johnson2019}. The median point is placed in red.}
    \label{fig:prospector_plot_corner}
        \end{center}
\end{figure*}

A parametric delayed-$\tau$ star formation history was used, with the timescale $\tau$ drawn from a log-uniform prior between $0.1$ and $10$~Gyr. The main stellar population parameters, including total mass $M$, age $t_{\mathrm{age}}$, e-folding time $\tau$, metallicity $\log Z/Z_{\odot}$, and dust attenuation $A_{V}$, were given broad uniform priors. A Chabrier initial mass function was assumed \citep{chabrier_galactic_2003} and dust attenuation was modeled with the Calzetti Milky Way dust law with $R_V=3.1$ \citep{calzetti_dust_1994}. Furthermore, we do not model various stellar emission lines. 
The galaxy's redshift was fixed to the spectroscopic value of $z = 0.084840$. The stellar continuum was supplemented with hot dust emission \citep{Draine2007} and with an AGN-heated dust component allowed to contribute a fraction $f_{AGN}$ of the total bolometric stellar luminosity of the galaxy \citep{Nenkova2008}. 
Given the posterior samples provided by \texttt{prospector}, we report the median galaxy properties listed in Table~\ref{tab:prospector}. Furthermore, for errors, we report the 16th and 84th percentile of the posterior sample. From the derived posteriors on $M$, $t_{\mathrm{age}}$ and $\tau$ we computed the stellar mass $M_*$, star formation rate (SFR) and mass-weighted stellar age $\mathrm{t_{m}}$ using the standard formulae \citep{Johnson2019}. 
The best-fit SED is shown in Figure \ref{fig:prospector_plot_SED} and a corner plot is shown in Figure \ref{fig:prospector_plot_corner}. The derived parameters are in general agreement to the host galaxy modeling presented by \citet{Gillanders25ulz}. 

\begin{table}
\centering
\caption{Median and 16th to 84th percentile of AT 2025ulz's host galaxy parameters based on our \texttt{prospector} modeling.}
\begin{tabular}{lcc}
\hline\hline
Parameter & Value\\
\hline
$\log(M_*/M_\odot)$      & $9.92^{+0.13}_{-0.15}$\\
$\log(Z/Z_\odot)$      & $-0.72^{+0.23}_{-0.25}$\\
$A_V$ [mag]            & $0.32^{+0.09}_{-0.10}$\\
$t_{\mathrm{age}}$ [Gyr]              & $6.04^{+3.96}_{-2.74}$\\
$\tau$ [Gyr]           & $1.23^{+0.88}_{-0.63}$\\
$\log(f_{\rm AGN})$    & $-1.58^{+0.22}_{-0.34}$\\
$\mathrm{SFR}$ [$M_\odot$/yr] & $0.52^{+0.15}_{-0.16}$\\
$\mathrm{t_{m}}$ [Gyr] & $3.83^{+2.36}_{-1.66}$\\
\hline
\label{tab:prospector}
\end{tabular}
\end{table}

\begin{figure}
    \centering
    \includegraphics[width=1\linewidth]{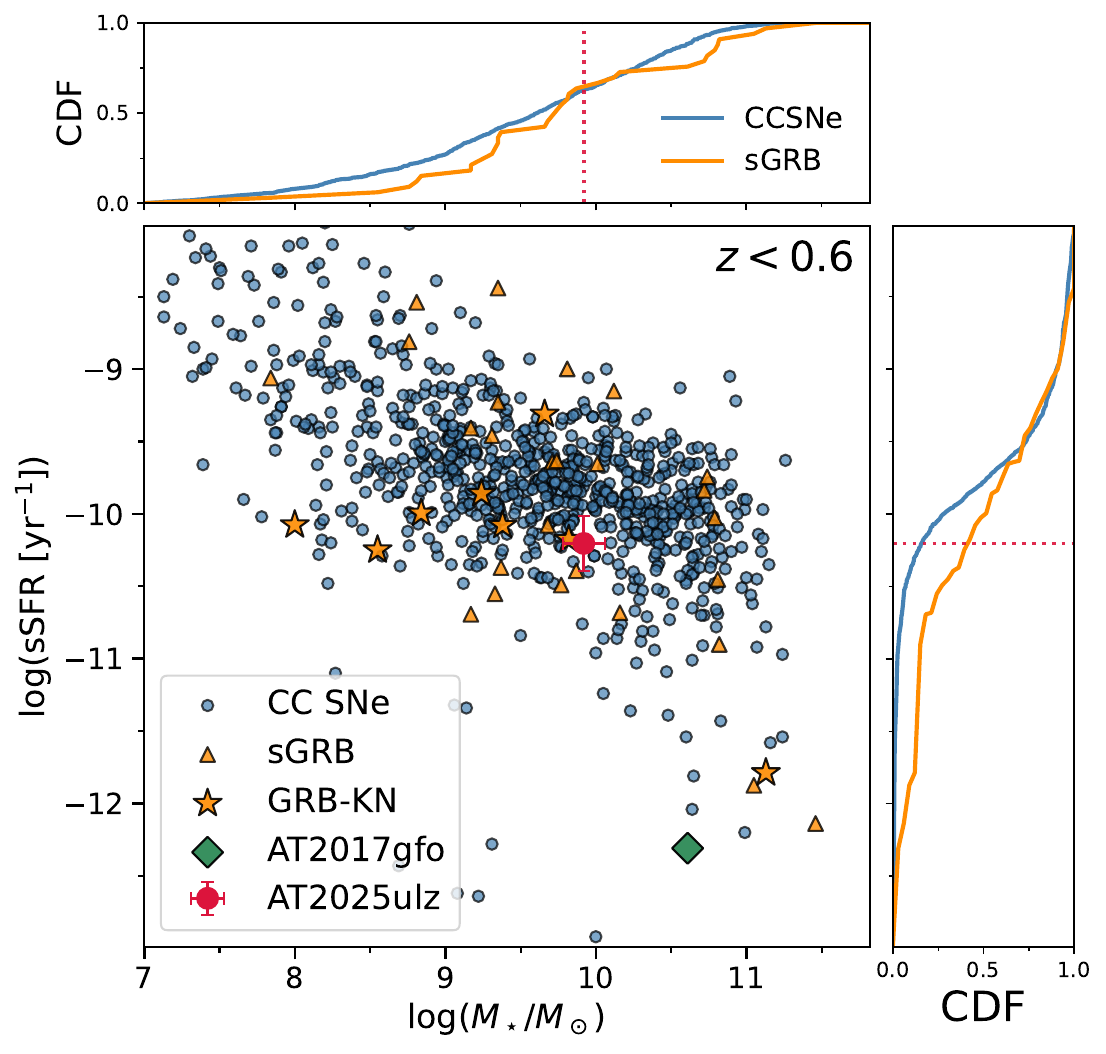}
    \caption{A comparison of the specific star formation rate (sSFR) versus stellar mass of CC SNe \citep{schulze_palomar_2021}, short GRBs \citep{nugent_short_2022,OConnor2022}, GRB-KN \citep{gal-yam_novel_2006,Jin2016,Rastinejad2022,Troja2022,Levan2023,yang_lanthanide-rich_2024}, and AT2017gfo \citep{palmese_evidence_2017,nugent_short_2022}, and AT2025ulz (see Table \ref{tab:prospector}). We have limited the redshift of all events shown here to $z<0.6$ for comparison to the highest-$z$ GRB-KN candidate \citep[GRB 200522A at $z=0.554$;][]{Fong2021,OConnor2021}. The position of AT2025ulz is consistent with both CC SNe and short GRBs.
    }
    \label{fig:massvssfr}
\end{figure}

\section{Comparison of the host galaxy properties to short GRB and core-collapse supernova hosts}

According to the BPT analysis (Figure \ref{fig:bpt}), the host of AT2025ulz appears to fall squarely within the star forming regime. This is in line with the fact that there is minimal AGN contribution found by the \texttt{prospector} fit (Table~\ref{tab:prospector}, with an AGN fraction of ${\sim}0.021$). The data here then suggest that the spectrum of the host of AT2025ulz is dominated by ongoing star formation rather than nuclear activity, consistent with what is typically observed in galaxies hosting core collapse supernovae and short gamma ray bursts. 

This is noticeably different from the host of GW170817 \citep[NGC 4993; see Figure 3 of][]{palmese_evidence_2017}. The galaxy there was markedly different, being an early-type system with an old stellar population and low star formation activity. Perhaps most uniquely, the host galaxy of GW170817 also had clear evidence of a recent galactic merger that led to the BNS that later coalesced in a KN event being flung far away outside of the galaxy. The list of merger signatures included disturbed stellar light profiles and kinematic peculiarities in observations. Such a merger event certainly had influenced the distribution of compact binary objects in this galaxy, a mechanism that would be missing in the case of the host of AT2025ulz. If an association is concretely demonstrated between AT2025ulz and S250818k, it would demonstrate an entirely different formation pathway than in the case of GW170817. In particular, it would potentially be of a far faster timescale formation channel associated to recent star formation.

Theoretical models have suggested that the formation of subsolar mass neutron star mergers in the accretion disk of a collapsar is possible \citep{PiroPfahl2007, Metzger2024, lerner_fragmentation_2025,ChenMetzger2025}. As AT2025ulz was suggested to be linked to both binary neutron star mergers and Type IIb supernovae \citep{Kasliwal2025sn}, we compare the host galaxy properties to the host galaxies of core-collapse supernovae \citep[CC SNe;][]{schulze_palomar_2021} and short duration gamma-ray bursts \citep[sGRB;][]{nugent_short_2022,OConnor2022}. We additionally highlight short GRBs with candidate kilonova associations, referred to as GRB-KN \citep{gal-yam_novel_2006,Jin2016,Berger2013,Tanvir2013,Troja150101B,Troja2019b,Fong2021,OConnor2021,Rastinejad2022,Troja2022,Levan2023,yang_lanthanide-rich_2024}. We limited our comparison to galaxies at $z<0.6$. Figure \ref{fig:massvssfr} shows a comparison between stellar mass and specific star formation rate (sSFR), which places AT2025ulz at a typical location in either population. 

We note that the SFR inferred from SED modeling is higher (by a factor of $\sim$\,1.35), but also within $2\sigma$, compared to the value inferred from the H$\alpha$ emission line ($0.386 \pm 0.104$ $M_\odot$/yr) even after correcting for the extinction inferred from the Balmer decrement (see Table \ref{tab:fastspecfit_output}). Additionally, the radio inferred SFR ($2.07 \pm 0.26$ $M_\odot$/yr; \citealt{2025GCN.41500....1B}) is higher than that (by a factor of $\sim$4), suggesting dust-obscured star formation in the host galaxy. This is reasonable to consider as the galaxy is found to be quite dusty from both emission line diagnostics and the SED modeling. Both the H$\alpha$ and radio SFR are inferred following \citet{murphy_calibrating_2011}. It is also possible that this difference is due to mismatches between the size of the DESI fiber (which does not capture the entire galaxy) and the size of the extended stellar population of the host galaxy, whereas both the host galaxy photometry and radio emission can capture the full extent (size) of ongoing star formation. Additionally, radio emission traces star formation on longer timescales ${\sim}100$ Myr compared to the shorter timescale ${\sim} 10$ Myr traced by H$\alpha$ \citep[e.g.,][]{Condon1992,2012ARA&A..50..531K,2013seg..book..419C}. Therefore, the deviation in the optically infered SFR and radio inferred SFR may be either due to dust-obscuration (above what is derived from the Balmer decrement) or a post-starburst episode with rapidly declining SFR in the last ${\sim}100$ Myr.

\section{Host Subtraction}

\begin{figure*}
    
    \centering
    \includegraphics[width=0.98\linewidth]{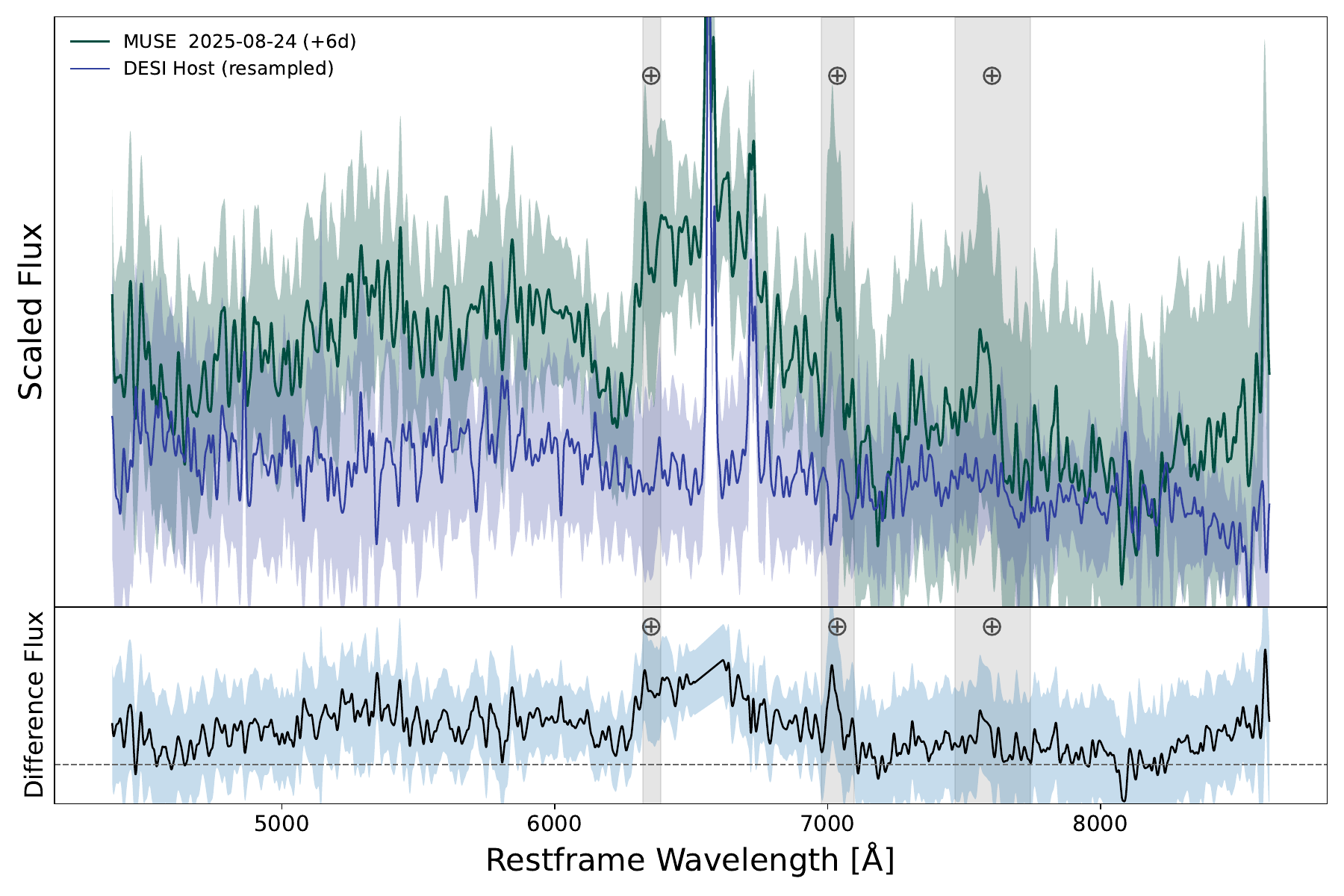}
    \caption{MUSE spectrum of AT2025ulz reported by \citet{banerjee_engrave_2025} and DESI host spectrum of AT2025ulz. The MUSE spectrum classifies AT2025ulz as a Type IIb supernova. The top panel shows both spectra normalized by their median flux while the bottom panel shows the MUSE spectrum with the DESI spectrum subtracted. Tellurics are denoted with $\oplus$ and the H$\alpha$ and NII complex is masked out in the subtraction. A clear P Cygni H$\alpha$ shape can be seen.}
    \label{fig:hostsub}
\end{figure*}

Host subtraction is a difficult but important task in transient astronomy. While host subtraction in photometry is relatively straightforward, spectroscopic host subtractions is difficult, also due to a lack of templates. Even with sufficient templates, instrumental issues with different resolution, aperture, slit losses, telluric corrections, and flux calibrations cause further difficulty. Thus proposed is a relatively naive but seemingly effective method to highlight broad lines in various spectra. The processes involves convolving spectra to a common resolution, then re-binning with interpolation to a common wavelength bin. In this case, the DESI spectrum is rebinned to the spectrum taken by MUSE \citep{banerjee_engrave_2025,2025GCN.41532....1B} approximately six days after the gravitational wave event. After binning, the spectra are normalized by the median flux over the common wavelength grid and then subtracted. The results of such a subtraction can be seen in Figure~\ref{fig:hostsub}. In both the original and subtracted, we can see the presence of a P Cygni H$\alpha$ line (as noted in \citealt{banerjee_engrave_2025}), which is a tell tale sign of a type II or IIb SN \citep[see also][]{Gillanders25ulz,Yang2025ulz,Franz2025}. Host subtraction is important, if this event was a KN, host subtraction could elucidate the heavy metal lines that indicate of R-processing and would show a tell-tale sign of a KN event.

\section{DESI Redshifts of Candidates}

\subsection{Probability of association to S250818k given the AT 2025ulz host redshift}

To assess whether AT2025ulz is potentially associated with the gravitational-wave candidate S250818k, we compute the probability that its host galaxy lies within the three-dimensional localization volume of the event. Following the formalism described in~\cite{Ashton:2017ykh}, we use the latest and publicly available \texttt{BILBY}~\citep{Ashton:2018jfp} localization skymap~\citep{Singer_2016} encoding the posterior density over sky position and luminosity distance, which can be factorized as
\begin{equation}
\label{eq:p3d}
p(\alpha,\delta,d_L) = p(d_L \mid \alpha,\delta)\, p(\alpha,\delta),
\end{equation}
where $p(\alpha,\delta)$ is the marginalized sky probability and $p(d_L \mid \alpha,\delta)$ is the conditional distance posterior along each line of sight.
The odds of association are given by the overlap integral defined as,
\begin{equation}
    \mathcal{I} = \int \frac{p(d_L(z_{\rm{EM}}) \mid \alpha_{\rm{EM}},\delta_{\rm{EM}})p(\alpha_{\rm{EM}},\delta_{\rm{EM}})}{\pi(\alpha_{\rm{EM}},\delta_{\rm{EM}})} .
\end{equation}
where $\pi(\alpha_{\rm{EM}},\delta_{\rm{EM}})$ is the probability for the position of the transient to have originated from a random association and where $(\alpha_{\rm{EM}},\delta_{\rm{EM}}, z_{\rm{EM}}) $ correspond to the AT2025ulz host galaxy position. 
We find $\log_{10}\mathcal{I} \approx 3.9-4.2$ depending on the value of the Hubble constant we use. For comparison purposes, the overlap integral for GW170817 and GRB 170817A had  $\log_{10}\mathcal{I} \sim 6$~\citep{Ashton:2017ykh,Piotrzkowski:2021hhy}.
To interpret this odds, we compare it against the following null hypotheses: a galaxy is drawn uniformly in comoving volume and uniformly on the sky with no distance information. The uniform-in-comoving volume prior corresponds to the expectation for a random galaxy population, while the uniform-sky prior corresponds to the chance alignment of a transient with the two-dimensional localization region. The ratio of the measured overlap to these null expectations provides an odds ratio for association versus random chance of coincidence. However, as emphasized in prior work~\citep{Ashton:2020kyr,MaganaHernandez:2024yfg,2024ApJ...977..122B}, spatial and distance consistency are necessary but not sufficient conditions for association: a robust claim requires a physical model that explains the transient’s observed evolution in the context of a compact binary merger.

\subsection{Other candidates}
Based on the tentative association of AT2025ulz to the GW candidate, we also report here DESI's contribution to the follow up of S250818k for other identified transients. We perform a cross-match of all candidates that were discovered after the GW event and positionally coincident transients to S250818k reported publicly on the transient name server (TNS) to the current DESI catalog as of 2025-09-26. In order to perform this crossmatch, first we find the three most probable hosts of any given transient using the projected distance from the host nucleus. We identify the most probable host in the Legacy Survey DR9 \citep{Dey2019AJ} using the Directional Light Radius (DLR) method \citep{Sullivan2006_4DLR,SmithDES_4DLR}. From this, we retain only convincing sources in Legacy Survey DR9 for host identification that satisfy two criteria: (1) robust detections in at least two of the $grz$ bands with $g < 24$, $r < 23.4$, and $z < 22.5$; and (2) physically plausible colors, restricted to $-1 < g - r < 4$. This permits the search of transients that are potentially very off-nuclear and broadens the set of potential sources.

There have been 152 transients reported to TNS within the 99\% confidence region of S250818k\footnote{\href{https://www.wis-tns.org/ligo/o4/S250818k_20250818_012006}{\url{https://www.wis-tns.org/ligo/o4/S250818k_20250818_012006}}}. We find DESI redshifts for 73 of these transients. We inspect the 15 for which the matches are within the volume, i.e.~the DESI redshift is consistent and find that 13 out of 15 have correctly matched hosts as indicated by an eye inspection of if the transient seemed best match to the DESI host and are within the volume. We compile the following candidates with redshifts that are within the $3\sigma$ space given Planck cosmology \citep{aghanim_planck_2020} in Table~\ref{tab:crossmatch}. Candidates that are outside the volume can be rejected as unrelated to the gravitational wave event. Those 58 transient candidates that are outside the reported volume can be found in the appendix in Table~\ref{tab:desi_match_bad}. 

\section{Targeted Follow-up of DESI Selected Hosts}

A third strategy in which DESI spectroscopic redshifts guide follow-up efforts of gravitational wave events is that they enable the identification of host galaxies whose three-dimensional position best matches the gravitational wave localization volume. With the coverage of DESI spectroscopy, particularly from the Bright Galaxy Survey, wide field spectroscopy  enables pinpointing the redshifts of the majority of the stellar mass in the range where BNS events are detectable by LVK. Targeted, deep follow-up with instruments that have only a limited field of view can then cover the actual host with substantial probability in the case of well-localized gravitational wave events.

We utilized this strategy, too, in the case of S250818k. From the latest catalog of DESI redshifts, we selected those 20 galaxies in the northern hemisphere part of the LVK localization region of S250818k that maximized the product of three-dimensional probability density and stellar mass (see Appendix \ref{ap:desi_hosts} for more details). In this, we marginalized over the uncertainty in cosmological parameters when converting redshift to luminosity distance (based on a broad uniform prior between the SH0ES \citep{riess_comprehensive_2022} and Planck \citep{aghanim_planck_2020} Hubble constant constraints), and used Legacy Survey $z$ band flux as a stellar mass proxy.

Follow-up of these candidate hosts commenced at 2025-08-19T21:50. Observations were conducted with the Fraunhofer Telescope at Wendelstein Observatory (FTW; \citealt{2014SPIE.9145E..2DH}) using the Three Channel Imager (3KK, \citealt{2016SPIE.9908E..44L}). We simultaneously imaged the $r$, $z$, and $J$ bands for $3 \times 180$~s on each of the first 10 potential host galaxies selected as described above, ordered by observability. The list of targets observed is provided in Appendix~B.

The images were reduced using a custom pipeline \citep[see][for details]{2002A&A...381.1095G, 2025A&A...701A.225B}. Astrometry was calibrated by matching to the Gaia catalog \citep{GAIA_DR3_2022}. We used the Pan-STARRS DR1 catalog \citep{2010SPIE.7733E..0EK} to determine zeropoints of the $r$ and $z$ band images, and the 2MASS catalog \citep{Skrutskie2006} for $J$ band.  

At the time of observations, 1.85 d post the gravitational wave event, a hypothetical kilonova would have been close to the peak of its red-optical emission. At the typical photometric depths achieved for points sources, $r\approx23.3$, $z\approx22.0$, $J\approx21.5$ mag ($3\sigma$, AB), such a transient would be detectable at high significance according to both models and Afterglow-KNe in the literature (e.g.~\citealt{Hall2025sn}). The field of view of the instrument is also suitable for this search: At the distance of $d_L = 259$ Mpc, the 7' FOV of the 3KK instrument, results in physical distances of more than 200 kpc from the center being observed, easily covering the full range of offsets of sGRBs from galactic centers \citep{Fong_2022,OConnor2022}.

We performed difference imaging of the $r$ and $z$ bands relative to Legacy Survey imaging. Visual inspection of the images does not reveal any candidate transient in our data. We repeated difference imaging analysis in $r$, $z$, and $J$ bands against template images we took ourselves on October 2, 2025, after the decay of a hypothetical transient. Here, too, no transient was detected.

\section{Conclusion}

In this work, we provide the properties of the host of the transient AT2025ulz,  hypothesized to be a candidate counterpart to the gravitational wave event S250818k \citep{Kasliwal2025sn,Hall2025sn}. We also reduce the list of other potential counterparts, based on DESI spectroscopic redshifts that place their hosts outside the gravitational wave event localization, and present the follow up carried out around DESI potential host galaxies with the 2.1m Fraunhofer Telescope at Wendelstein Observatory. 

One novel advantage of DESI’s extensive redshift catalog is the ability to rapidly identify the most promising transients for further observations during gravitational wave events follow-up campaigns. Having an immediate, precise redshift reference allows observers to quickly reject candidates that lie outside the localization volume or that exhibit absolute magnitudes inconsistent with a kilonova. For the case of AT 2025ulz, we measure an integral overlap of $\log_{10}\mathcal{I} \approx 3.9-4.2$, which, based on the 3D location of this transient with respect to the GW candidate, provides some compelling evidence towards association and against chance coincidence, although this result does not account for the physical interpretation of the transient emission as arising from the GW source. This capability significantly improves the efficiency of follow-up efforts by narrowing the field to likely counterparts. In the case of S250818k alone, hundreds of transients were discovered within the sky localization region \citep{Kasliwal2025sn,Gillanders25ulz,Franz2025}. Roughly half of these already had a redshift associated with their likely respective host galaxy in DESI. A crucial component of gravitational-wave follow-up is not only the identification of the true counterpart, but also the confident rejection of all other candidates. Since it is impractical to maintain rigorous follow-up of every transient, DESI’s redshift catalog represents a major advance in our ability to prioritize and eliminate candidates effectively.

Another critical use of DESI is the ability to perform spectroscopic host subtraction, revealing the intrinsic transient spectrum and minimizing host contamination to feature identification. As demonstrated (Figure \ref{fig:hostsub}), host subtraction is required for detecting both broad features found in spectra as well as determining the continuum color of transient spectra. Further work into building a more rigorous method to use DESI spectra to produce host subtracted spectra is warranted. Such an approach could involve using time concurrent and historic photometry to better constrain both continuum and transient properties. Galactic properties vary over the face of the entire galaxy, by taking photometry at a fiber size aperture compared to the nucleus (where the DESI spectrum is taken) one could recalibrate the continuum and then when subtracting use difference photometry to calibrate the subtracted spectrum. Such rigor in spectral subtractions will prove essential for the discovery of these intrinsically faint compact object mergers. If in the future more candidates are found within bright galaxies, it will be equally essential to remove the contaminating galaxy light.

Finally, DESI offers a new method of surveying for telescopes mounted without wide field survey instruments. Dedicated observations of in-volume candidate host galaxies offer a new potential path for a smaller class of telescopes in the era the Vera C. Rubin observatory. As next generation gravitational wave detectors come online, the distance of discovered compact mergers will be out or reach of any wide field survey smaller than Rubin. However, through deep observations of high-probability hosts, we will be able to better leverage smaller telescopes to still find these faint and distant events.

With the end of LIGO O4c approaching and the imminent onset of the Vera C. Rubin Observatory Legacy Survey of Space and Time (LSST), it is crucial to continue efforts to uncover these rare but fascinating compact binary mergers. Leveraging galaxy catalogs to conduct rigorous searches for orphan kilonovae will be essential for building a robust sample that can be further expanded with the future LIGO-Virgo-KAGRA O5 or A\# events. 

\begin{acknowledgments}

This work is supported by NSF Grant No. 2308193. B. O. is supported by the McWilliams Postdoctoral Fellowship in the McWilliams Center for Cosmology and Astrophysics at Carnegie Mellon University. This material is based upon work supported by the U.S. Department of Energy (DOE), Office of Science, Office of High-Energy Physics, under Contract No. DE–AC02–05CH11231, and by the National Energy Research Scientific Computing Center, a DOE Office of Science User Facility under the same contract. Additional support for DESI was provided by the U.S. National Science Foundation (NSF), Division of Astronomical Sciences under Contract No. AST-0950945 to the NSF’s National Optical-Infrared Astronomy Research Laboratory; the Science and Technology Facilities Council of the United Kingdom; the Gordon and Betty Moore Foundation; the Heising-Simons Foundation; the French Alternative Energies and Atomic Energy Commission (CEA); the National Council of Humanities, Science and Technology of Mexico (CONAHCYT); the Ministry of Science, Innovation and Universities of Spain (MICIU/AEI/10.13039/501100011033), and by the DESI Member Institutions: \url{https://www.desi.lbl.gov/collaborating-institutions}. Any opinions, findings, and conclusions or recommendations expressed in this material are those of the author(s) and do not necessarily reflect the views of the U. S. National Science Foundation, the U. S. Department of Energy, or any of the listed funding agencies.

The authors are honored to be permitted to conduct scientific research on I'oligam Du'ag (Kitt Peak), a mountain with particular significance to the Tohono O’odham Nation.

This paper contains data obtained at the Wendelstein Observatory of the Ludwig-Maximilians University Munich.
Funded by the Deutsche Forschungsgemeinschaft (DFG, German Research Foundation) under Germany's Excellence Strategy – EXC-2094 – 390783311.

This work used resources on the Vera Cluster at the Pittsburgh Supercomputing Center (PSC). Vera is a dedicated cluster for the McWilliams Center for Cosmology and Astrophysics at Carnegie Mellon University. We thank the PSC staff for their support of the Vera Cluster.

This publication makes use of data products from the Wide-field Infrared Survey Explorer, which is a joint project of the University of California, Los Angeles, and the Jet Propulsion Laboratory/California Institute of Technology, funded by the National Aeronautics and Space Administration.

This publication makes use of data products from the Two Micron All Sky Survey, which is a joint project of the University of Massachusetts and the Infrared Processing and Analysis Center/California Institute of Technology, funded by the National Aeronautics and Space Administration and the National Science Foundation.

\end{acknowledgments}

\facilities{CTIO:Mayall, Wendelstein:FTW}



\appendix

\section{DESI Candidates}

Here we list two tables. The first table is a list of all transient reporeted to TNS within the localization of the LVK alert and have redshifts consistent with such a volume. The second table includes transient candidates that are rejected due to a host that is outside of the 3$\sigma$ probability region provided by LVK for S250818k.

\begin{table*}[ht!]
\centering
\caption{Cross-match of publicly reported transients temporally and positionally coincident with S250818k to the DESI catalog (as of 2025-09-26). We offer the publically avaible classification of a transient if it is avaible.}
\label{tab:crossmatch}
\begin{tabular}{lcccccl}
\hline
TNS Name & RA host & Dec host & $z$ & $D_{L}$ [Mpc] & Discovery & Note \\
\hline
2025uxs & 236.0596 & 27.4798 & 0.06275 & 291.1 & \citet{chambers_pan-starrs_2025_08_22} & Ia-SN; \citep{wise_ztf_2025}  \\
2025utx & 251.0454 & 42.4654 & 0.08009 & 376.0 & \citet{chambers_pan-starrs_2025_08_21} & \\
2025vnt & 237.5920 & 29.0245 & 0.07551 & 353.4 & \citet{sollerman_ztf_2025} & Ia-SN; \citep{fremling_ztf_2025} \\
2025uzd & 251.2640 & 44.2982 & 0.06733 & 313.3 & \citet{stein_ztf_2025} & \\
2025utr & 271.9428 & 56.8697 & 0.05592 & 258.2 & \citet{chambers_pan-starrs_2025_08_21} & \\
2025utq & 262.9458 & 54.7942 & 0.08100 & 380.5 & \citet{chambers_pan-starrs_2025_08_21} & \\ 
2025usy & 268.8042 & 56.0342 & 0.08574 & 404.1 & \citet{chambers_pan-starrs_2025_08_21} & \\ 
2025uqe & 258.0304 & 49.2303 & 0.07303 & 341.2 & \citet{stein_ztf_2025} & \\
2025upw & 222.0821 & 9.1172  & 0.02921 & 132.3 & \citet{tonry_atlas_2025} & \\
2025ulz & 237.9757 & 30.9026 & 0.08484 & 399.6 & \citet{GCN41414} & II-SN; \citep{2025GCN.41532....1B} \\
2025uso & 244.2394 & 35.0560 & 0.02517 & 113.6 & \citet{chambers_pan-starrs_2025_08_21} & IIb-SN; \citep{sollerman_ztf_2025_09_29}\\
2025unu & 239.5424 & 42.9910 & 0.06114 & 283.3 & \citet{jones_yse_2025} & \\
2025wcl & 36.4880  & 1.1867  & 0.07291 & 340.6 & \citet{ofek_last_2025} & \\
\hline
\end{tabular}
\end{table*}

\begin{longtable}{lccccc}
\caption{Cross-match of transients with DESI catalog entries whose redshift confidently places them outside the GW localization region.} \label{tab:desi_match_bad} \\

\toprule
TNS Name & RA host [deg] & Dec host [deg] & $z$ & Spectype & $D_{L}$ [Mpc] \\
\midrule
\endfirsthead

\toprule
TNS Name & RA host & Dec host & $z$ & Spectype & $D_{L}$ [Mpc] \\
\midrule
\endhead

\bottomrule
\endfoot
 2025uus & 244.2946 &  39.6485 & 0.1479 &   GALAXY &      725.8524 \\
 2025uuq & 244.6472 &  37.9424 & 0.1214 &      QSO &      586.0073 \\
 2025uul & 237.0139 &  29.7992 & 0.4025 &      QSO &     2256.6342 \\
 2025vaj & 245.1992 &  42.0982 & 0.2295 &   GALAXY &     1180.8826 \\
 2025uzl & 241.6002 &  37.7827 & 0.2003 &   GALAXY &     1013.7973 \\
 2025vaa & 254.1146 &  46.8251 & 0.6117 &   GALAXY &     3721.9598 \\
 2025uzx & 242.0437 &  38.3653 & 0.4523 &      QSO &     2590.5359 \\
 2025wek & 237.0823 &  32.7603 & 0.7406 &   GALAXY &     4696.9613 \\
 2025vjj & 236.3173 &  29.5315 & 0.1115 &   GALAXY &      534.7696 \\
 2025uvu & 236.7140 &  29.9974 & 0.1131 &   GALAXY &      542.6445 \\
 2025uvt & 238.8827 &  31.0152 & 0.6878 &   GALAXY &     4291.7750 \\
 2025uuc & 264.9935 &  53.3381 & 2.6640 &      QSO &    22550.7862 \\
 2025uua & 261.4404 &  51.9287 & 0.2819 &      QSO &     1492.0705 \\
 2025uxo & 248.8887 &  41.0160 & 0.1327 &   GALAXY &      645.0393 \\
 2025uvs & 269.2136 &  55.3507 & 0.1918 &   GALAXY &      966.3277 \\
 2025utz & 247.7828 &  40.7122 & 0.2576 &   GALAXY &     1346.3430 \\
 2025usn & 237.6129 &  30.3335 & 0.2167 &   GALAXY &     1107.2874 \\
 2025vyh & 272.3377 &  59.0274 & 1.2842 &   GALAXY &     9243.8722 \\
 2025vqh & 270.4048 &  59.0388 & 0.1501 &   GALAXY &      737.4811 \\
 2025vpk &  43.2179 & -11.1208 & 0.1109 &   GALAXY &      531.5151 \\
 2025vki & 289.0243 &  61.5418 & 0.1703 &      QSO &      846.8556 \\
 2025vad & 237.2357 &  28.6868 & 0.2982 &      QSO &     1591.4464 \\
 2025vax & 241.0787 &  32.3593 & 0.3191 &   GALAXY &     1720.6413 \\
 2025vau & 234.3892 &  29.6584 & 0.5778 &   GALAXY &     3474.1519 \\
 2025vat & 233.0135 &  27.5936 & 0.8388 &   GALAXY &     5471.1382 \\
 2025uog &  58.3271 & -33.1842 & 0.3482 &   GALAXY &     1904.6131 \\
 2025unn & 239.1649 &  33.2565 & 0.3183 &   GALAXY &     1715.6246 \\
 2025unm & 247.0437 &  42.0718 & 0.3651 &   GALAXY &     2012.6763 \\
 2025unl & 250.1464 &  46.7457 & 0.2294 &   GALAXY &     1180.3280 \\
 2025wge & 242.0833 &  40.0865 & 0.1323 &   GALAXY &      642.7407 \\
 2025uzn & 235.2263 &  31.3966 & 0.2921 &   GALAXY &     1553.9217 \\
 2025vap & 265.4473 &  52.2285 & 0.1716 &   GALAXY &      854.4513 \\
 2025vae & 234.2501 &  24.4684 & 0.2061 &   GALAXY &     1046.8841 \\
 2025uuk & 262.3959 &  54.6045 & 0.2089 &   GALAXY &     1062.8393 \\
 2025uui & 239.7301 &  30.9933 & 0.1736 &   GALAXY &      865.0194 \\
 2025uuh & 283.7587 &  60.1951 & 0.1645 &   GALAXY &      815.3669 \\
 2025uzg & 233.7462 &  24.8625 & 0.2292 &   GALAXY &     1179.1234 \\
 2025uzf & 255.1250 &  49.6714 & 0.1688 &   GALAXY &      839.0211 \\
 2025unk & 246.3895 &  40.7262 & 0.4073 &   GALAXY &     2288.6124 \\
 2025wku & 282.0333 &  60.8339 & 0.1943 &   GALAXY &      980.0172 \\
 2025utt & 267.5892 &  54.9656 & 0.2685 &      QSO &     1411.0327 \\
 2025uql & 267.2358 &  57.8760 & 0.1478 &   GALAXY &      725.1096 \\
 2025ung & 251.2007 &  44.2291 & 0.1803 &      QSO &      902.0566 \\
 2025wnj & 257.3367 &  54.1327 & 0.1663 &   GALAXY &      825.3396 \\
 2025wmh &  43.5906 &  -6.5685 & 0.1402 &   GALAXY &      684.6676 \\
 2025wiq &  35.6168 &   3.1288 & 0.2159 &   GALAXY &     1102.4901 \\
 2025vak & 237.6181 &  38.5757 & 0.5645 &      QSO &     3377.9328 \\
 2025vah & 239.4819 &  29.2198 & 1.2998 &   GALAXY &     9382.7814 \\
 2025uzt & 236.2475 &  35.7668 & 0.2820 &   GALAXY &     1492.4074 \\
 2025uph & 297.0632 &  64.6738 & 0.1814 &   GALAXY &      908.4743 \\
 2025unr & 220.8747 &  13.1996 & 0.1189 &   GALAXY &      572.5861 \\
 2025vtg &  37.8268 &  -0.1797 & 0.2751 &   GALAXY &     1450.6962 \\
 2025vmd &  39.4021 &  -9.5487 & 0.1580 &   GALAXY &      780.1118 \\
 2025wcz &  35.7978 &   1.9576 & 0.2057 &   GALAXY &     1044.5975 \\
 2025vvs &  37.6256 &  -0.6878 & 0.2932 &   GALAXY &     1560.6515 \\
 2025vun &  37.6251 &   3.0939 & 0.2454 &   GALAXY &     1274.0367 \\
 2025wew &  36.9475 &   1.1849 & 1.9112 &      QSO &    15065.3336 \\
 2025wdi &  35.0614 &   1.9105 & 0.1234 &   GALAXY &      595.9829 \\
\end{longtable}

\section{Targeted follow-up of DESI candidate host galaxies}
\label{ap:desi_hosts}

Here we list the DESI selected candidate hosts followed up with 3KK/FTW. The target selection was based on a combination of the individual 3D probability P3D (similar to \citealt{Singer_2016}) of each galaxy w.r.t.~the gravitational wave localization as well as the luminosity $L_r$ in $r$-band as a proxy for stellar mass. Potential targets were drawn from a daily updated DESI catalog with application of common quality cuts. At the time of observations, only the \texttt{bayestar.multiorder.fits,1} map was available. The product (P3D $\cdot L_r$) is unitless after being normalized to sum to one in the limit that galaxies follow a Schechter luminosity function ($\Phi_* = 1.0\cdot 10^{-9} h^{-3}\text{Mpc}^{-3}\text{dex}^{-1}$, $M_* - 5\log(h) = 21.0$, $\alpha = 1.27$) and are randomly distributed in 3D. Therefore, this quantity represents an estimate of the probability that each target is the true host. The total probability covered by these ten galaxies sums up to  $\approx $1\%. A Kilonova-like transient would have been confidently detected in the $r$-band for all targets when comparing with AT2025gfo. In $i$-band and $J$-band, only S250818k\_162333\_p390716, S250818k\_161052\_p415309 and S250818k\_160742\_p385551 would have been detected. The absolute magnitudes of the other targets in these bands were slightly fainter than the reference values from AT2025gfo. The absolute magnitude reference values of AT2025gfo, taken 1.927, 1.863 and 2.170 days post merger in $r$-, $i$- and $J$-bands are -14.55, -14.83 \citep{Andreoni2017} and -15.32 \citep{Utsumi_2017, Villar_2017}.

\begin{longtable}{lccccccccc}
\caption{List of potential host galaxies identified by their 3D localization including DESI redshifts followed up with targeted 3KK/FTW observations, based on the full \texttt{bayestar.multiorder.fits,1} map. No transients were seen in difference imaging. The table includes the internal name, Right Ascension and Declination in degrees, host redshift $z$, luminosity distance $D_L$, normalized true host probability P3D $\cdot L_r$, the time difference between the gravitational wave signal $t_0$ and start of observations $t_{\text{obs}}$ and the individual 3$\sigma$ absolute magnitude limit in the respective bands $M^{\text{lim}}_{r/i/J}$} \label{tab:desi_ftw} \\

\toprule
Internal name & RA [deg] & Dec [deg] & $z$ & $D_L$ [Mpc] & P3D$\cdot L_r$ & $t_{\text{obs}} - t_0 $ [d] & $M^{\text{lim}}_r$ & $M^{\text{lim}}_i$ & $M^{\text{lim}}_J$ \\
\midrule
\endfirsthead

\toprule
Internal name & RA [deg] & Dec [deg] & $z$ & $D_L$ [Mpc] & P3D$\cdot L_r$ & $t_{\text{obs}} - t_0 $ [d] & $M^{\text{lim}}_r$ & $M^{\text{lim}}_i$ & $M^{\text{lim}}_J$ \\
\midrule
\endhead

\bottomrule
\endfoot

S250818k\_160742\_p385551 & 241.9235 & 38.9307 & 0.03851 & 175.6 & 0.00103 & 1.8324 & -12.76 & -14.02 & -15.05 \\
S250818k\_162333\_p390716 & 245.8864 & 39.1211 & 0.03503 & 159.3 & 0.00100 & 1.8398 & -12.70 & -13.97 & -15.24 \\
S250818k\_160011\_p400610 & 240.0455 & 40.1027 & 0.06199 & 287.4 & 0.00071 & 1.8472 & -13.95 & -15.18 & -15.56 \\
S250818k\_155915\_p400626 & 239.8124 & 40.1073 & 0.07070 & 329.8 & 0.00067 & 1.8547 & -14.33 & -15.56 & -15.91 \\
S250818k\_161800\_p412332 & 244.5018 & 41.3922 & 0.06120 & 283.6 & 0.00072 & 1.8621 & -13.86 & -15.12 & -15.39 \\
S250818k\_161052\_p415309 & 242.7175 & 41.8858 & 0.03826 & 174.4 & 0.00084 & 1.8695 & -12.90 & -14.14 & -14.46 \\
S250818k\_161833\_p430453 & 244.6389 & 43.0814 & 0.05992 & 277.4 & 0.00200 & 1.8769 & -13.88 & -15.13 & -15.49 \\
S250818k\_161505\_p433544 & 243.7694 & 43.5955 & 0.06516 & 302.8 & 0.00155 & 1.8843 & -14.11 & -15.45 & -15.79 \\
S250818k\_162631\_p435826 & 246.6295 & 43.9740 & 0.05917 & 273.8 & 0.00081 & 1.8917 & -13.89 & -15.22 & -15.57 \\
S250818k\_170818\_p452118 & 257.0755 & 45.3551 & 0.06057 & 280.6 & 0.00077 & 1.8992 & -13.87 & -15.26 & -15.59 \\
\end{longtable}


\bibliography{sample701}{}
\bibliographystyle{aasjournalv7}



\end{document}